\documentclass[%
reprint,superscriptaddress,
 amsmath,amssymb,
 aps,
 prl,
]{revtex4-2}

\usepackage{amsmath}
\usepackage{graphicx}
\usepackage{dcolumn}
\usepackage{bm}
\usepackage[normalem]{ulem}
\usepackage{color}
\usepackage{array}
\usepackage{microtype}
\usepackage[colorlinks=false, allcolors=blue]{hyperref}
\usepackage{etoolbox}

\makeatletter
\let\orig@bibinfo\bibinfo
\renewcommand{\bibinfo}[2]{%
  \ifdefstring{#1}{doi}{}{%
    \orig@bibinfo{#1}{#2}%
  }%
}
\makeatother

\begin{document}
\title{JAX-in-Cell: A Differentiable Particle-in-Cell Code for Plasma Physics Applications}
\author{Longyu Ma}
\email{lma95@wisc.edu}
\author{Rogerio Jorge} 
\author{Hongke Lu}
\author{Aaron Tran}
\author{Christopher Woolford}
\affiliation{Department of Physics, University of Wisconsin--Madison, Madison, Wisconsin 53562, USA}

\begin{abstract}

JAX-in-Cell is a fully electromagnetic, multispecies, and relativistic 1D3V Particle-in-Cell (PIC) framework implemented entirely in JAX. It provides a modern, Python-based alternative to traditional PIC frameworks. It leverages Just-In-Time compilation and automatic vectorization to achieve the performance of traditional compiled codes on CPUs, GPUs, and TPUs. The resulting framework bridges the gap between educational scripts and production codes, providing a testbed for differentiable physics and AI integration that enables end-to-end gradient-based optimization. The code solves the Vlasov-Maxwell system on a staggered Yee lattice with either periodic, reflective, or absorbing boundary conditions, allowing both an explicit Boris solver and an implicit Crank-Nicolson method via Picard iteration to ensure energy conservation. Here, we detail the numerical methods employed, validate against standard benchmarks, and showcase the use of its auto-differentiation capabilities.

\end{abstract}

\maketitle

\section{Statement of Need}

A plasma is a collection of free ions and electrons whose self-generated
and external electromagnetic fields drive collective behavior. Such behavior can be modelled using PIC simulations, which offer a fully
kinetic description and enable exploration of complex interactions in
modern plasma physics research, such as fusion devices, laser-plasma interactions, and astrophysical plasmas \cite{birdsall1991plasma}.
However, such simulations can be computationally very expensive, often requiring hardware-specific implementations in low-level languages.
The current landscape of high-performance production codes such as OSIRIS \cite{Fonseca2002}, EPOCH \cite{Smith2021}, VPIC \cite{Le2023}, and WARPX \cite{WarpX}, which are written in C++ or Fortran with MPI/CUDA backends, are highly optimized for massively parallel simulations, often with complex compilation chains, and require external adjoint implementations for optimization. This imposes a steep barrier to entry for new developers, making it cumbersome to test new algorithms, as well as to integrate with modern data science workflows.
On the other hand, open-source educational Python scripts typically lack the performance and capabilities needed to perform cutting-edge research.

JAX-in-Cell is able to fill this gap by implementing a 1D3V PIC framework entirely within the JAX ecosystem \cite{jax2018github}. It is open-source, user-friendly, and developer-friendly, written entirely in Python. It addresses three specific needs not met by existing codes: 1) hardware-agnostic high performance; 2) unified explicit and implicit solvers; 3) differentiable physics and AI integration. This is achieved using JAX's Just-In-Time (JIT) compilation via XLA, which allows us to achieve performance parity with compiled languages on CPUs, GPUs, and TPUs. Therefore, researchers can prototype new algorithms in Python and immediately use them on complex situations and accelerated hardware. Furthermore, JAX-in-Cell is inherently differentiable due to its automatic differentiation (AD) capabilities. This allows for new research directions such as optimization of laser pulse shapes, parameter discovery from experimental data, or embedding PIC simulations in Physics-Informed Neural Network loops. Finally, both an explicit scheme using the standard Boris algorithm \cite{boris1970relativistic} and a fully implicit, energy-conserving scheme \cite{CHEN20142391} are available to cross-verify results and perform long simulations with large time steps, a capability often lacking in lightweight tools.

\section{Structure}

%
As in standard PIC approaches \cite{birdsall1991plasma}, in JAX-in-Cell, particles are advanced along characteristics of the Vlasov equation
\begin{equation}
\partial_t f_s + \mathbf{v}\cdot\nabla f_s
+ \frac{q_s}{m_s} \left( \mathbf{E} + \mathbf{v} \times \mathbf{B} \right)
\cdot \nabla_{\mathbf{u}} f_s = 0,
\end{equation}
with the electromagnetic fields governed by the standard Maxwell equations. Here, $\mathbf{v}$ is velocity, $q_s$ and $m_s$ are the
particle charge and mass, $\mathbf{E}$ and $\mathbf{B}$ are the
electric and magnetic fields, $\mathbf{u} = \mathbf{v}\gamma$ is the
proper velocity, and $\gamma = \sqrt{1 + u^2/c^2}$ is the Lorentz
factor, with $c$ the speed of light.
The distribution function $f_s$ is discretized as
\begin{equation}
f_s(\mathbf{x}, \mathbf{v}) \approx \sum_{p} w_p\, \delta(\mathbf{x} - \mathbf{x_p})\, \delta(\mathbf{v} - \mathbf{v_p}),
\end{equation}
where $x_p$ denotes the position of each pseudo-particle, $\mathbf{v_p}$ denotes the velocity of each pseudo-particle, the weight is given by $w_p = n_0L/N$, with
$n_0$ number density, $L$ the spatial domain length and $N$ the number of pseudo-particles for that species. Then, the spatial domain is divided into $N_x$ uniform cells with spacing $\Delta x$ and advanced in time by $\Delta t$. To mitigate numerical noise, each pseudo-particle is represented by a triangular shape function spanning three grid cells, and the same kernel is used consistently for both the
particle-to-grid charge deposition and the grid-to-particle field interpolation \cite{hockney1988computer}. Accordingly, the current density $\mathbf J$ is computed from the continuity equation using a discretely charge-conserving scheme \cite{villasenor1992rigorous} consistent with the shape function.
%

\begin{figure}
\centering
\includegraphics[width=\linewidth, trim={2.7cm 3.85cm 3.9cm 3.95cm},clip]{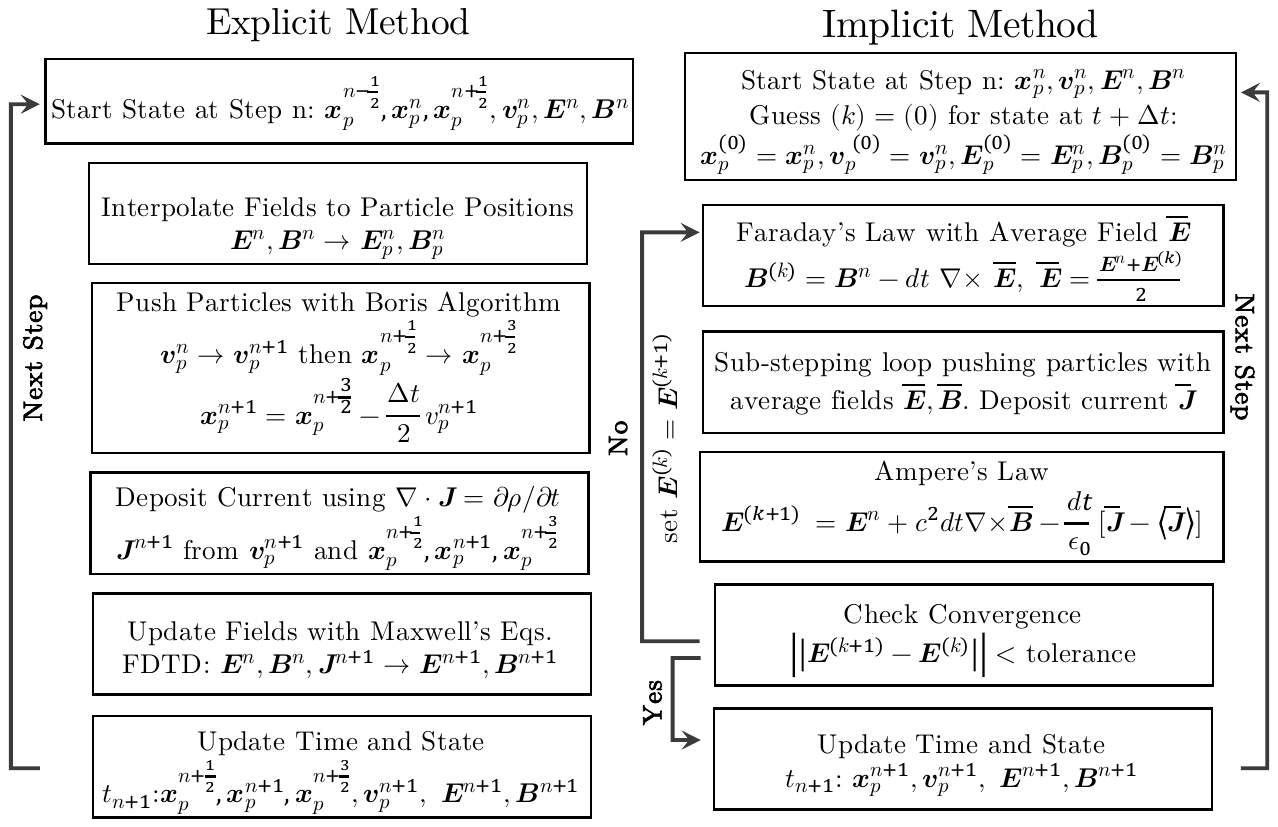}
\caption{Time-stepping algorithms in JAX-in-Cell. Left: explicit Boris time-stepper and a Finite-Difference Time-Domain (FDTD) method using a staggered Yee grid for the electromagnetic fields. Right: implicit Crank-Nicolson time stepper using a Picard iteration for the electromagnetic system.
\label{fig:algorithm}}
\end{figure}

The core logic of JAX-in-Cell is distributed along six specific modules. The first one is \texttt{simulation.py}, which serves as the high-level entry point, handling parameter initialization (via TOML parsing), memory allocation for particle and field arrays, and the execution of the main loop. The time-stepping is performed using a JAX primitive \texttt{jax.lax.scan}, which allows it to be optimized by the XLA compiler. Then, \texttt{algorithms.py} implements the time integration schemes, which advance the system state by one timestep $\Delta t$ by a sequence of operations, namely particle push, source deposition, and field update. We implement two time-evolution methods (see Fig.~\ref{fig:algorithm}), an explicit Boris algorithm \cite{boris1970relativistic}, and an implicit Crank--Nicolson scheme solved via Picard iteration \cite{CHEN20142391}. The JIT-compiled particle dynamics is present in \texttt{particles.py}, which includes the relativistic and non-relativistic Boris rotation and the field interpolation routines, which are heavily vectorized using \texttt{jax.vmap}. The electromagnetic solvers are present in \texttt{fields.py}, which include the finite-difference curl operators for the Faraday and Ampère laws, as well as the divergence cleaning routines that enforce charge conservation. The deposition of charge and current densities from particle positions onto the grid is handled by \texttt{sources.py}, which implements high-order spline interpolation and digital filtering to mitigate aliasing noise. Finally, \texttt{boundary\_conditions.py} is the centralized module to enforce boundary constraints, including routines for particle reflection/absorption and ghost-cell updates for the electromagnetic fields.

Due to its simplified design, JAX-in-Cell is able to pass the entire simulation state between functions, which is maintained as an immutable tuple referred to in the code as the \texttt{carry}$=(\mathbf E, \mathbf B, \mathbf x_p, \mathbf v_p, q, m)$.
This allows the entire simulation to be treated as a single differentiable function, which can facilitate the integration of automatic differentiation workflows.
In order to reduce kernel launch overheads on GPUs, as well as the vector lengths for different populations, JAX-in-Cell adopts a monolithic array formulation of the multi-species architecture optimized for the Single-Instruction-Multiple-Data (SIMD) paradigm of JAX.
That is, the simulation state, \texttt{carry}, is a single concatenated state of unified, global arrays, regardless of the number of physical species defined in the input configuration. This is a different approach than the one used in traditional PIC codes, which employ an object-oriented design with different species stored in separate containers and iterated over sequentially. During initialization, the code parses the TOML configuration file for the list of species and, for each population, the function \texttt{make\_particles} generates each phase-space distribution and weights at the initial time, which are then appended to the global state vectors. This allows load balancing across GPU cores and minimizes warp divergence.
A comprehensive set of \texttt{pytest}-based unit and integration tests is included in JAX-in-Cell.
The test suite is part of a continuous integration workflow on GitHub actions across multiple Python and JAX versions, and uploads coverage reports to Codecov.



\section{Capabilities}

%

The electromagnetic fields are defined on a staggered Yee lattice. The code then solves the Ampere and Faraday equations
\begin{equation}
    \frac{\partial \mathbf E}{\partial t}=c^2 \nabla \times \mathbf B - \frac{\mathbf J}{\epsilon_0},~ \frac{\partial \mathbf B}{\partial t} = - \nabla \times \mathbf E,
\end{equation}
which, in the 1D geometry ($\partial/\partial y = \partial/\partial z = 0$), reduces the curl operator to spatial derivatives along the $x$ lattice. The discrete spatial derivative operator uses central differences for internal points, with ghost cells handling the boundary conditions (periodic, reflective, or absorbing). JAX-in-Cell allows flexible boundary conditions by handling particle trajectories and electromagnetic fields independently at the domain edges.
Charge conservation is ensured using a divergence cleaning step, where the longitudinal component $E_x$ is projected to satisfy Gauss' law, $\nabla \cdot \mathbf E = \rho/\epsilon_0$.
The charge density $\rho$ and current density $\mathbf J$ are deposited onto the grid using a quadratic (3-point) spline shape function $S(x)$, where a multi-pass binomial digital filter is applied to the source terms to mitigate grid-heating instability. This effectively suppresses high wavenumbers near the Nyquist limit while preserving the macroscopic (low wave-number) plasma dynamics.

While the user can initialize their own distribution functions, JAX-in-Cell allows users to use a predefined perturbed Maxwellian
\begin{equation}
f_s(x, \mathbf{v}, t = 0) = f_{s0}(\mathbf{v})\left[ 1 + a \cos\left(\frac{2 \pi k x}{L}\right)\right],
\end{equation}
where $a$ is the perturbation amplitude and $k$ the perturbation
wavenumber. The Maxwellian background $f_{s0}$ is given by the following anisotropic, drifting distribution
\begin{equation}
\begin{aligned}
f_{s0}(\mathbf{v}) &= \frac{n_0}{2 \, \pi^{3/2} v^3} 
\left[ 
e^{- \frac{(v_x - v_{bx})^2}{v_{th,x}^2} - \frac{(v_y - v_{by})^2}{v_{th,y}^2} - \frac{(v_z - v_{bz})^2}{v_{th,z}^2}}\right. \\
&\quad + 
\left.e^{- \frac{(v_x + v_{bx})^2}{v_{th,x}^2} - \frac{(v_y + v_{by})^2}{v_{th,y}^2} - \frac{(v_z + v_{bz})^2}{v_{th,z}^2}}
\right],
\end{aligned}
\end{equation}
with $v^3 = v_{th,x} v_{th,y} v_{th,z}$ the product of each thermal velocity along $(x,y,z)$ directions, $v_{bi}$ the drift velocities along each direction $i$, and $n_0$ the background density. The addition of an opposite sign drift velocity $\pm v_b$ is controlled using the \texttt{velocity\_plus\_minus} input parameter.
Such a perturbed Maxwellian allows us to perform many benchmark simulations, such as Landau damping, the two-stream instability, the bump-on-tail instability, and the Weibel instability.

\begin{figure}
\centering
\includegraphics[width=\linewidth]{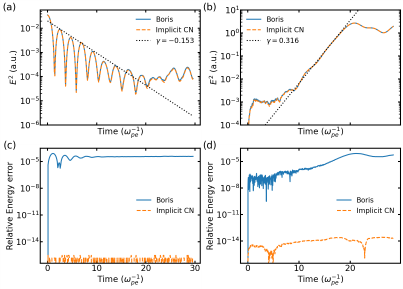}
\caption{Electric field energy evolution for Landau damping and the two-stream instability. (a) Landau damping with analytical damping rate $\gamma = 0.153\omega_{pe}$. (b) Two-stream instability showing fitted exponential growth rate. (c--d) Relative total energy deviation $|E_{\text{total}} - E_{\text{total}}(0)| / E_{\text{total}}(0)$ demonstrating energy conservation.\label{fig:output}}
\end{figure}

We first validate JAX-in-Cell by performing such simulations in a periodic boundary of length $L$, and compare with the corresponding linear theory. Linearizing the Vlasov--Maxwell system around this initial distribution (with equal thermal velocities $v_{th,x}=v_{th,y}=v_{th,z}$), yields the dispersion relation
\begin{equation}
1 + \frac{1}{2k^2\lambda_D^2}
\left[  2 + \xi_1 Z(\xi_1)+\xi_2 Z(\xi_2)\right] = 0, \quad
\xi_i=\frac{\omega}{kv_{th}}-\frac{v_{b_i}}{v_{th}},
\end{equation}
where $\lambda_D$ is Debye length and $Z$ is the Fried--Conte plasma dispersion function. The complex frequency $\omega$ determines both the oscillation frequency and the damping or growth rate $\gamma$.
We use such a theoretical model to demonstrate two representative test cases using non-relativistic algorithms. 1) Landau damping using a perturbation: $a = 0.025$,
$k\lambda_D = 1/2$, zero drift velocities $v_{b_x} = v_{b_y} = v_{b_z} = 0$,
$v_{th} = 0.35\,c$, resolution of $N = 40000$ particles, $N_x = 32$ grid cells, $\Delta x = 0.4\lambda_D$ gird spacing and $\Delta t = 0.1\,\omega_{pe}^{-1}$ timestep; 2) two-stream instability using a perturbation $a = 5\times10^{-7}$,
$k\lambda_D = 1/8$, velocities $v_{b_x} = 0.2\,c$ with \texttt{velocity\_plus\_minus} set to true, $v_{th} = 0.05\,c$, $N = 10000$, $N_x = 100$, $\Delta x = 0.5 \lambda_D$ and $\Delta t = 0.1\,\omega_{pe}^{-1}$
We show in Fig.~\ref{fig:output} (top) the evolution of the electric field energy, as well as the fitted damping/growth rates, showing good agreement with the analytical prediction. We also show in Fig.~\ref{fig:output} (bottom) the resulting relative energy error between the explicit and implicit methods to demonstrate the precision of the implicit method.

\begin{figure}
\centering
\includegraphics[width=\linewidth]{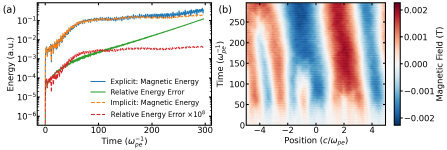}
\caption{Weibel instability. (a) Evolution of the magnetic field energy and the relative energy error of the simulation during the Weibel instability. (b) Spatial profile of the magnetic field $B_y$.\label{fig:Weibel}}
\end{figure}

Next, we perform a simulation of the Weibel instability, which arises in anisotropic plasmas and may be a mechanism for magnetic field generation and application. The plasma is initialized with an anisotropic velocity distribution with $v_{thz}/v_{thx}=\sqrt{T_z/T_x}=10$. We show in Fig. \ref{fig:Weibel} the evolution of the magnetic field energy and the energy relative error (left) and the magnetic field strength in the $y$ direction (right). As expected, the magnetic field organizes into filamentary structures perpendicular to the velocity anisotropy. Initially, multiple small filaments form, which subsequently merge into larger-scale structures as the system evolves. In this case, the implicit method shows a bounded relative energy error of at most $10^{-11}$, while the explicit method appears to have unbounded energy errors.

\begin{figure}
\centering
\includegraphics[width=\linewidth]{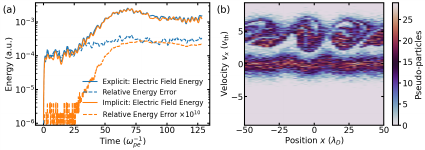}
\caption{Simulation of the bump-on-tail instability. The number of pseudo-particles in the bulk and beam populations is equal, with a beam-to-bulk weight ratio of $3\times10^{-2}$. (a) Time evolution of the electric field energy and relative energy error. (b) Snapshot of phase space at $80\,\omega_{pe}^{-1}$.
\label{fig:bump-on-tail}}
\end{figure}

We demonstrate the multi-species capability of JAX-in-Cell by performing a simulation of the bump-on-tail instability (Fig.~\ref{fig:bump-on-tail}). This instability arises when a high-velocity electron beam creates a positive slope in the electron velocity distribution. We initialize the plasma with a bulk Maxwellian electron population and a small, high-velocity beam that produces a pronounced bump in the tail of the distribution, and we track the evolution of the phase-space density and the associated electric field. During the linear growth phase, resonant electrons exchange energy with Langmuir waves, leading to exponential amplification of the electric field. As the instability evolves, the initially smooth electron distribution develops coherent phase-space structures, illustrating the code's ability to capture nonlinear wave--particle interactions.

\begin{figure}
\centering
\includegraphics[width=\linewidth]{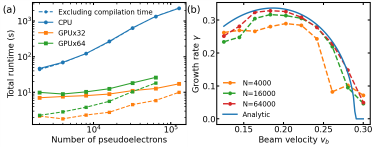}
\caption{(a) Comparison of total runtime between CPU and GPU. (b)
Influence of pseudo-particle number on the two-stream instability
sampling results. Growth rates computed from exponential
fits.\label{fig:Run_time}}
\end{figure}

Additionally, we evaluate the computational performance of the code by comparing CPU and GPU run times. For this test, we use the hardware present in the NERSC Perlmutter HPE Cray EX supercomputer, that is, an AMD EPYC 7763 CPU and an NVIDIA A100 GPU. We assess how the total runtime scales with the number of pseudo-particles. As a representative benchmark, we simulate ten drift velocities drawn from the two-stream dispersion relation. The results are shown in Fig.~\ref{fig:Run_time}. We find that, for the same workload, the GPU executes the simulation approximately two orders of magnitude faster than the CPU. In particular, for a system of 64000 pseudo-particles, the GPU completes the full drift-scan in about six seconds after the initial compilation. The benchmarks also indicate that GPU results depend on floating-point
precision: running in 32-bit mode by manually disabling JAX's x64 option reduces memory usage and improves speed, but can introduce deviations when compared to 64-bit results. 
%

\begin{figure}
    \centering
    \includegraphics[width=\linewidth]{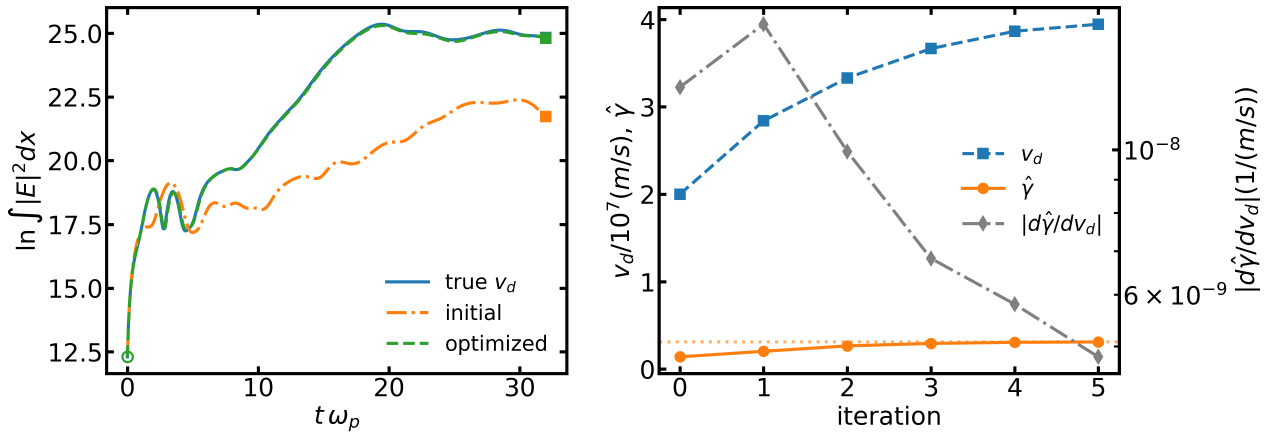}
    \caption{Demonstration of autodifferentiation capabilities in JAX-in-Cell for optimization using the two-stream instability. Left: time evolution of the electric field energy for the true drift speed $v_d$, the initial guess (dashed), and the optimized value. 
    Right: optimization history of the drift speed $v_d$, the dimensionless growth rate $\hat{\gamma}=\gamma/\omega_p$, and the auto-differentiated sensitivity $|d\hat{\gamma}/dv_d|$. The algorithm converges in six iterations.}
    \label{fig:two_stream_inverse}
\end{figure}

Finally, we showcase in Fig. \ref{fig:two_stream_inverse} one of the central motivations for JAX-in-Cell, that is, to make plasma physics simulations natively differentiable. This allows us to take gradients of high-level parameters used for diagnostics with respect to any number of physical or numerical parameters without resorting to finite differencing. Such a capability is increasingly important for tasks such as Bayesian inference of transport coefficients, real-time control, design of laser pulses, and training of hybrid physics-ML surrogates for fusion devices or astrophysical plasmas. However, we note that automatic differentiation in PIC codes might be challenging due to noise, integer indexing, and discontinuous shape functions. We show in Fig. \ref{fig:two_stream_inverse} how some of these challenges can be overcome. We optimize the dimensionless growth rate $\hat \gamma = \gamma/\omega_{pe}$ of the two-stream instability by applying a damped Newton update to the drift velocity $v_d$ using JAX to evaluate both $\hat \gamma(v_d)$ and its derivative $d\hat \gamma/d v_d$ via a single forward-mode Jacobian-vector product, which allows for efficient memory handling. The drift speed (squares) converges from $2 \times 10^7$ m/s to $3.95 \times 10^7$ m/s in six iterations, while $\hat \gamma$ (circles) rapidly approaches the target growth rate, and the sensitivity $|d \hat \gamma/dv_d|$ (diamonds) decreases as the optimizer approaches the solution.
%
Future work will involve generalizing such an optimization to inverse problems, such as the possibility of forward-modelling electron beams responsible for Type III solar radio bursts, and the design of new experiments.

\section{Acknowledgments}

This work was supported by the National Science Foundation under Grant No.~PHY-2409066. This work used the Jetstream2 at Indiana University through allocation PHY240054 from the Advanced Cyberinfrastructure Coordination Ecosystem: Services \& Support (ACCESS) program which is supported by National Science Foundation grants \#213859, \#2138286, \#2138307, \#2137603 and \#2138296. This research used resources of the National Energy Research Scientific Computing Center, a DOE Office of Science User Facility supported by the Office of Science of the U.S. Department of Energy under Contract No.~DE-AC02-05CH11231 using NERSC award NERSC DDR-ERCAP0030134. Aaron Tran was supported by the DOE Fusion Energy Sciences Postdoctoral Research Program, administered by the Oak Ridge Institute for Science and Education (ORISE) and Oak Ridge Associated Universities (ORAU) under DOE contract DE-SC0014664.

\bibliographystyle{apsrev4-2}
\bibliography{apssamp}

\end{document}